\documentclass[%
 reprint,
superscriptaddress,
 amsmath,amssymb,
 aps,
]{revtex4-2}

\usepackage{graphicx}
\usepackage[caption=false]{subfig}

\newcommand{\ie}{i.e., }
\newcommand{\eg}{e.g., }
\newcommand{\quotes}[1]{``#1''}

\begin{document}

\title{Accelerating the identification of informative reduced representations \\ of proteins with deep learning for graphs}

\author{Federico Errica}
\affiliation{Department of Computer Science, University of Pisa, Italy}
\author{Marco Giulini}
\affiliation{Physics Department, University of Trento, via Sommarive, 14 I-38123 Trento, Italy}
\affiliation{INFN-TIFPA, Trento Institute for Fundamental Physics and Applications, I-38123 Trento, Italy}
\author{Davide Bacciu}
\affiliation{Department of Computer Science, University of Pisa, Italy}
\author{Roberto Menichetti}
\affiliation{Physics Department, University of Trento, via Sommarive, 14 I-38123 Trento, Italy}
\affiliation{INFN-TIFPA, Trento Institute for Fundamental Physics and Applications, I-38123 Trento, Italy}
\author{Alessio Micheli}
\affiliation{Department of Computer Science, University of Pisa, Italy}
\author{Raffaello Potestio}
\affiliation{Physics Department, University of Trento, via Sommarive, 14 I-38123 Trento, Italy}
\affiliation{INFN-TIFPA, Trento Institute for Fundamental Physics and Applications, I-38123 Trento, Italy}

\date{\today}

\begin{abstract}
The limits of molecular dynamics (MD) simulations of macromolecules are steadily pushed forward by the relentless developments of computer architectures and algorithms. This explosion in the number and extent (in size and time) of MD trajectories induces the need of automated and transferable methods to rationalise the raw data and make quantitative sense out of them. Recently, an algorithmic approach was developed by some of us to identify the subset of a protein's atoms, or mapping, that enables the most informative description of it. This method relies on the computation, for a given reduced representation, of the associated mapping entropy, that is, a measure of the information loss due to the simplification. Albeit relatively straightforward, this calculation can be time consuming. Here, we describe the implementation of a deep learning approach aimed at accelerating the calculation of the mapping entropy. The method relies on deep graph networks, which provide extreme flexibility in the input format. We show that deep graph networks are accurate and remarkably efficient, with a speedup factor as large as $10^5$ with respect to the algorithmic computation of the mapping entropy. Applications of this method, which entails a great potential in the study of biomolecules when used to reconstruct its mapping entropy landscape, reach much farther than this, being the scheme easily transferable to the computation of arbitrary functions of a molecule's structure.
\end{abstract}

\maketitle

\section{Introduction}
\label{sec:intro}

Molecular dynamics (MD) simulations \cite{md_general_method, md_sim_biomol} are an essential and extremely powerful tool in the computer-aided investigation of matter. The usage of classical, all-atom simulations has boosted our understanding of a boundless variety of different physical systems, ranging from materials (metals, alloys, fluids, etc.) to biological macromolecules such as proteins. As of today, the latest software and hardware developments have pushed the size of systems that MD simulations can address to the millions of atoms \cite{SINGHAROY20191098}, and the time scale that a single run can cover can approach the millisecond for relatively small molecules \cite{shaw2009millisecond}.

\begin{figure}[htpb]
    \centering
    \includegraphics[width=\columnwidth]{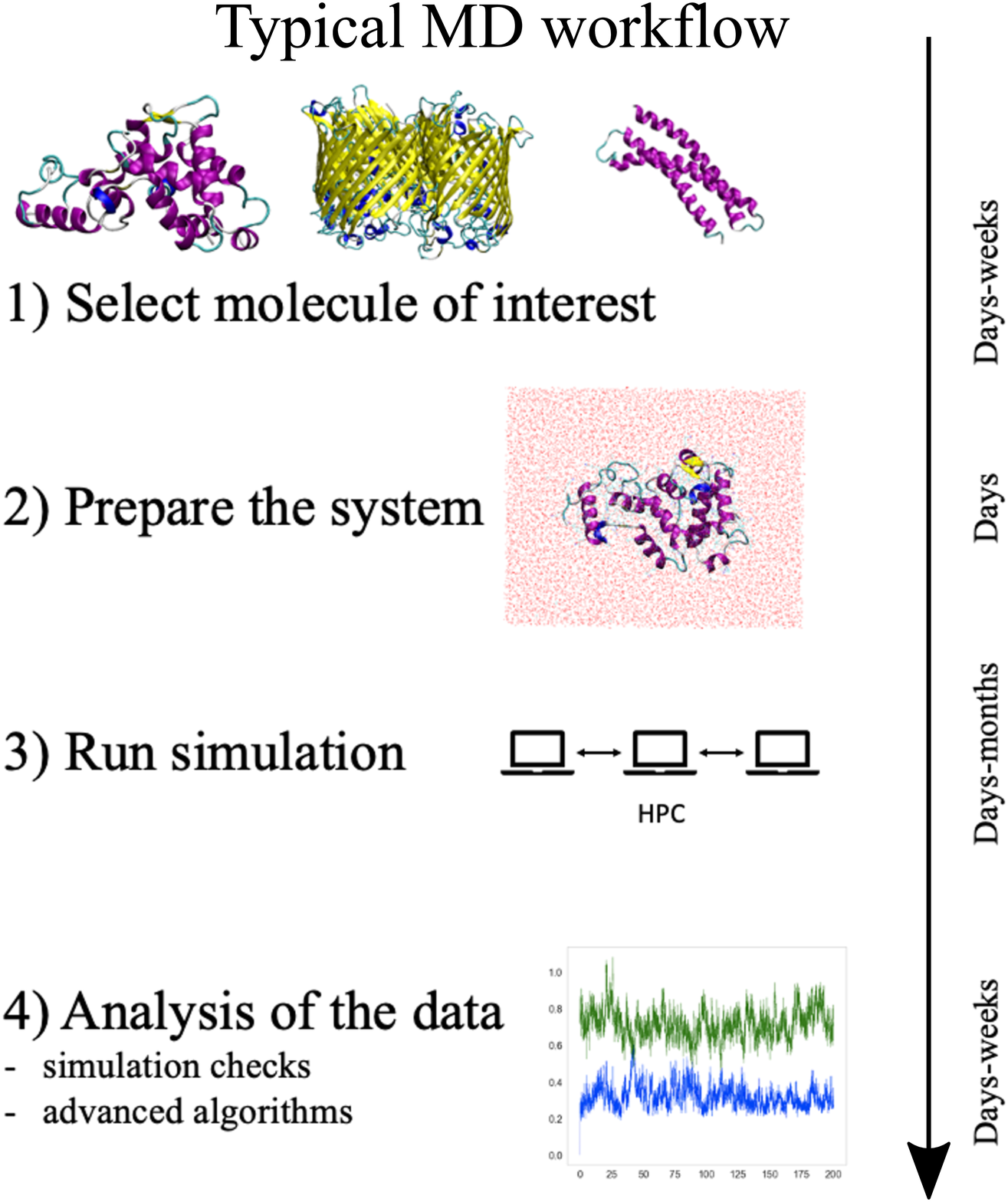}
    \caption{Schematic representation of the typical workflow of a molecular dynamics study. On the right we report the typical time scales required for each step of the process.}
    \label{fig:md_workflow}
\end{figure}

In general, a traditional molecular dynamics-based study proceeds in four steps, here schematically summarised in Figure \ref{fig:md_workflow}. First, the system of interest has to be identified; this apparently obvious problem can actually require a substantial effort {\it per se}, e.g. in the case of dataset-wide investigations. Second, the simulation has to be set up, which is another rather nontrivial step \cite{kandt2007setting}. Then it has to be run, typically on a high performance computing cluster. Finally, the simulation output has to be analysed and rationalised {\it in order to extract information from the data}.

This last step is particularly delicate, and it is acquiring an ever growing prominence as large and long simulations can be more and more effortlessly performed. The necessity thus emerges to devise a parameter-free, automated ``filter'' procedure to describe the system of interest in simpler, intelligible terms and make sense out of the immense amount of data we can produce - but not necessarily understand.

In physics, the most prominent example of a systematic, algorithmic modelling procedure is the renormalisation group \cite{ma2018modern, fisher1998renormalization}, a technique, developed for the study of critical phenomena, that relies on the iterative simplification of a detailed model, or \emph{coarse-graining} (CG). As the simplification progresses, the relevant properties of the system emerge while the irrelevant ones fade out. Coarse-graining methods have found vast application also outside of the realm of critical phenomena, specifically in the field of soft and biological matter \cite{marrink2007martini, Takada2012, Potestio2014, Saunders2013}. Here, they are employed to construct simplified representations of a given macromolecular system that have fewer degrees of freedom with respect to a reference, more detailed model, while retaining key features and properties of interest. In biophysical applications this amounts to describe a biomolecule, such as a protein, using a number of constituent units, called CG sites, lower than the number of atoms composing the original, all-atom system.

The coarse-graining process in soft matter requires two main ingredients, namely the representation and the CG potential. The former consists in the definition of a \emph{mapping} \cite{noid_mapping, rudzinski_2011, foley2015impact}, that is, the transformation ${\bf M}({\bf r}) = {\bf R}$ that connects a high-resolution description ${\bf r}$ of the system's configuration to a low-resolution one ${\bf R}$. The latter ingredient is the set of interactions the user introduces among the sites to reproduce the properties of the all-atom system.

During the last decades, substantial effort has been invested in the correct parameterisation of the CG potential \cite{noid_mapping, Shell2008}; on the other hand, the definition of the mapping has received much less attention, being selected by intuition in the vast majority of cases. A notable example of this natural choice for proteins is the $C_{\alpha}$ mapping, in which only the $\alpha$ carbon atoms of the polypeptide chain are retained. Although this choice can make the tuning of CG interactions easier, it is not always appropriate to account for the specific physico-chemical features of a protein.

Most methods developed in the field of soft matter do not make use of a system-specific, algorithmic procedure for the selection of the effective sites, but rather rely on general criteria, based on physical and chemical properties, to group together atoms in coarse-grained \quotes{beads} irrespectively of their local environment and global thermodynamics. While acceptable in most practical applications, this approach entails substantial limitations: in fact, the coarse-graining process implies a loss of information and, through the application of universal mapping strategies, system-specific properties, albeit relevant, might be \quotes{lost in translation} from a higher to a lower resolution representation. Hence, a method would be required that enables the automated identification of which subset of retained degrees of freedom of a given system preserves the largest amount of important detail from the reference, while at the same time reducing the complexity of the problem.

The challenge of the identification of a maximally informative mapping has been recently tackled by some of us \cite{giulini2020information}; specifically, we developed an algorithmic procedure to find a CG mapping that minimises the amount of information that is lost when the number of degrees of freedom with which one observes a system is decreased. The quantity that measures this loss is called mapping entropy \cite{foley2015impact,  rudzinski_2011, Shell2008, Shell2012}:
\begin{equation}
S_{map}=k_B\int d{\bf r}\ p_r({\bf r})\ln \left[ \frac{V^n}{V^N} \frac{p_r({\bf r})}{p_R({\bf M}({\bf r}))} \right]\geq0.
\label{eq:smap_general}
\end{equation}

Here, $p_r({\bf r})$ is the probability of sampling a configuration ${\bf r}$ in the high resolution system, namely the Boltzmann distribution $p_r({\bf r})\propto\exp(-\beta u({\bf r}))$, where $u({\bf r})$ is the  atomistic potential and $\beta = 1/k_B T$ is the inverse temperature. $p_R({\bf R})$ is the probability of sampling the configuration ${\bf R} = {\bf M}({\bf r})$ in the low resolution description:
\begin{equation}
\label{eq:pmacro}
 p_R({\bf R}) =\frac{1}{Z}\int d{\bf r}e^{-\beta u({\bf r})}\delta({\bf M}({\bf r}) - {\bf R}),
\end{equation}
where $Z$ is the canonical partition function of the system.

In Ref.~\cite{giulini2020information} and in this work we restrict our analysis to a peculiar functional form for the CG mapping, called \emph{decimation}, which consists in the selection of a subset of the system's $n$ atoms. These can thus be in one of two possible states, i.e. retained or not.
This choice is particularly appealing as it enables a rather straightforward bridging between molecular modelling and deep learning. In fact, a protein structure can be represented as a graph in which each atom corresponds to a vertex, and edges connect pairs of atoms closer than a selected threshold. At odds with other definitions of a CG site, such as the centre-of-mass for which the exact knowledge of the real space coordinates is required, the information about the decimation mapping can be directly encoded in the vertices of the protein's graph by using a mark, \eg a binary value as it is done in this work.

In the case of a decimation of the all-atom system, $S_{map}$ can be written using the following notation \cite{rudzinski_2011}:
\begin{eqnarray}
\label{eq:smap_main}
S_{map} &=&  k_B\times D_{KL}(p_{r}({\bf r})||\bar{p}_r({\bf r})) \nonumber \\
&=& k_B\int d{\bf r}\ p_r({\bf r}) \ln \left[ \frac{p_r({\bf r})}{\bar{p}_r({\bf r})} \right],
\end{eqnarray}
that is, a Kullback-Leibler divergence $D_{KL}$ \cite{kullback1951information} between $p_r({\bf r})$ and the distribution $\bar{p}_r({\bf r})$ obtained by observing the former through the ``coarse-graining grid'', i.e., in terms of the selected CG mapping. $\bar{p}_r({\bf r})$ is defined as \cite{rudzinski_2011}
\begin{equation}
\label{eq:pbar}
\bar{p}_r({\bf r}) = {p_R({\bf M}({\bf r}))}/{\Omega_1({\bf M}({\bf r}))},
\end{equation}
where
\begin{equation}
\label{eq:omega1}
\Omega_1({\bf R}) =   \int d{\bf r}\  \delta({\bf M}({\bf r}) - {\bf R}),
\end{equation}
is the number of microstates ${\bf r}$ that map onto the CG configuration ${\bf R}$.

Different choices of CG mapping lead to different $\bar{p}_r({\bf r})$ and, consequently, to different values of $S_{map}$. Unfortunately, the definitions in Eq.~\ref{eq:smap_general} or \ref{eq:smap_main} do not allow, given a CG representation, to directly calculate the associated mapping entropy.

In Ref.~\cite{giulini2020information} Giulini {\it et al.} tackled this problem by establishing a connection between Eq. \ref{eq:smap_main} and the energetics of the system:
\begin{equation}
\label{eq:smap_expansion}
S_{map} \simeq k_B \int d{\bf R}\ p_R({\bf R}) \frac{\beta^2}{2}\langle(U-\langle U\rangle_{\beta|{\bf R}})^2\rangle_{\beta|{\bf R}},
\end{equation}
where $\langle U\rangle_{\beta|{\bf R}}$ is the average microscopic energy restricted to macrostate ${\bf R}$,
\begin{equation}
\langle U\rangle_{\beta|{\bf R}}=\int dU P_{\beta}(U|{\bf R})U,    
\end{equation}
and
\begin{eqnarray}
&&P_{\beta}(U|{\bf R})=\frac{p_R(U,{\bf R})}{p_R({\bf R})} \nonumber \\
&&=\frac{1}{p_R({\bf R})}\int d{\bf r}\ p_r({\bf r})\delta({\bf M}({\bf r}) - {\bf R})\delta(u({\bf r})-U)
\end{eqnarray}
is the conditional probability for the system to have energy $U$ provided that it is in macrostate ${\bf R}$.

Eq.~\ref{eq:smap_expansion} highlights that the mapping entropy can be calculated as a weighted average over all CG macrostates ${\bf R}$ of the variances of the atomistic potential energies of all configurations ${\bf r}$ that map onto a specific macrostate. Importantly, this allows one to measure $S_{map}$ given a set of all-atom configurations and a decimation mapping ${\bf M}$. Specifically, in Ref.~\cite{giulini2020information} this was achieved by lumping together configurations based on a distance matrix, the latter obtained by considering only the atoms retained in the mapping. Then, the energy variances over these clusters were computed, each one weighted with its probability.

The following, natural step in this analysis is then to identify the reduced representations of a system that are able to preserve the maximum amount of information from the all-atom reference, i.e., that minimise the mapping entropy.

However, for a molecule with $n$ atoms the number of possible decimation mappings is $2^n$, an astronomical amount even for the smallest proteins (with $n \sim 50$ one has more than $10^{15}$ distinct mappings). Even restricting the analysis to a fixed number of retained degrees of freedom (atoms) $N$, the number is still huge, namely ${n!}/(N! \ (n-N) !)$. Hence, in Ref.~\cite{giulini2020information} we resorted to a stochastic exploration of this immense space. In particular, given a CG mapping with randomly chosen atoms, we started adding/removing atoms with an acceptance criterion based on a Metropolis rule, using $S_{map}$ as loss function. By 
repeating this procedure for 48 independent runs, we built a set of optimised CG mappings such that their value of $S_{map}$ was a local minimum. Noteworthy, these mappings were found to more likely retain atoms directly related to the biological function of the proteins of interest, thus linking the described information-theoretical approach to the properties of biological systems. Hence, the protocol developed in Ref.~\cite{giulini2020information} represents not only a practical way to select the most informative mapping of a molecule, but also a promising paradigm to employ CGing as a controllable filtering procedure that can highlight relevant regions in a biological structure. 

The downside of this approach is its non-negligible computational cost, which is due to two factors:

\begin{enumerate}

    \item Eq. \ref{eq:smap_expansion} requires a set of configurations of the high-resolution system that are sampled through an MD simulation, which can be expensive from a computational perspective;
    
    \item the stochastic exploration of the set of possible CG mappings is limited and time-consuming.
    
\end{enumerate}

Table \ref{tab:time} shows a few values of the computational time required to perform these steps. It is worth stressing that the proteins studied here are small; therefore, these values would dramatically increase in the case of bigger biomolecules.

\begin{table*}
\centering
\begin{tabular}{@{}lllll@{}}
\toprule
Protein &  \textit{MD CPU time} & \textit{MD walltime} & \textit{Single measure} & \textit{Opt. time}\\ \hline
tamapin (PDB code 6d93) & $40.7$ days  & $2.55$ days & $\simeq 2.1 $ mins  & $\simeq 1.0 $ days \\
 adenylate kinase (PDB code 4ake) & $153.9$ days & $3.20$ days   & $\simeq 8.0 $ mins & $\simeq 1.2 $ days\\  \hline
\end{tabular}
\caption{\label{tab:time}An example of the computational cost of detailed simulations and mapping optimisation. Specifically, \textit{MD CPU time} (\textit{MD walltime}) represents the core time (user time) necessary to run 200 ns of molecular simulation on GROMACS 2018 package \cite{van2005gromacs}. \textit{Single measure} is the amount of time that is required to compute $S_{map}$ of a mapping. \textit{Opt. time} quantifies the computational burden of a single stochastic optimization ($2 \times 10^4$ steps): this value is not $\text{\textit{Single measure}} \times 2 \times 10^4$, since in \cite{giulini2020information} we resort to substantial approximations to speed-up the overall process.}
\end{table*}

The ultimate aim of this work is the development of a machine learning approach to predict the value of $S_{map}$ for a CG mapping associated with a protein. The first nontrivial requirement one can ask the model is to to reproduce the values of $S_{map}$ from a single biological structure for which we already have the ground truth. The model would not eliminate the need for steps $1$ and $2$, but it would make an extensive exploration of a protein's mapping space feasible. To remove these steps altogether, the model should also be able to transfer its knowledge to unknown proteins for which we do not have an all-atom simulation or data on CG mappings.We do not address this latter issue in the present work, but rather we focus on the development and assessment of a machine learning model for protein-specific $S_{map}$ prediction.
 
\begin{figure}[t]
    \centering
    \includegraphics[width=\columnwidth]{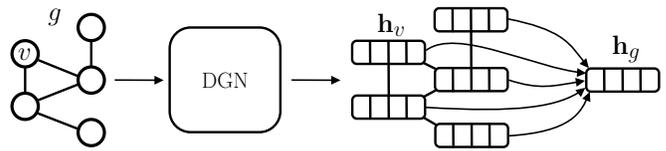}
    \caption{High-level overview of typical deep learning methodologies for graphs. A graph $g$ is given as input to a Deep Graph Network, which outputs one vector, also called embedding or state, for each vertex $v$ of the graph. In this work, we aggregate all those embeddings to obtain a single one that encodes the whole graph structure. Then, this is used by a standard machine learning regression algorithm to output the $S_{map}$ value associated with $g$.}
    \label{fig:dgn}
\end{figure}

With their long and successful story both in the field of coarse-graining \cite{gfeller2007spectral, depabloJCTC2019} and in the prediction of protein properties \cite{borgwardt2005protein, fout2017protein, torng2019graph}, graph-based algorithms represent the most natural choice to tackle such challenges. Here, we do not use graphs to develop a novel coarse-graining strategy, but rather we show that a graph-based machine learning algorithm reproduces the results of a coarse-graining procedure obtained by means of a lengthy and non-trivial optimisation process. Importantly, our approach makes use of a negligible fraction of the huge amount of information employed in the original method, i.e., a single protein structure viewed as a graph.

To predict $S_{map}$ values given a protein and a CG mapping, we leverage Deep Graph Networks (DGNs) \cite{bacciu_gentle_2020, micheli_neural_2009, scarselli_graph_2009}, a family of neural networks that naturally deal with relational data. The main advantages of DGNs are the efficiency and the ability of learning from graphs of different size and shape. This is possible for two reasons: first, DGNs focus on local processing of vertex neighbors, so computation can be easily parallelized on every graph; secondly, in a way that is similar to Convolutional Neural Networks \cite{lecun_convolutional_1995} for images, DGNs stack multiple layers of graph convolutions to propagate information. At each layer $\ell$, DGNs compute the state of each vertex $v$ as a vector $\mathbf{h}_v^{\ell}$ of fixed dimension. Eventually, a DGN produces a final embedding for each vertex that we call $\mathbf{h}_v$ as well as a global graph embedding $\mathbf{h}_g$. The latter is obtained by simple aggregation of all vertex embeddings as sketched in Figure \ref{fig:dgn}. Being in vectorial form, $\mathbf{h}_g$ can be fed into standard machine learning algorithms to solve graph regression/classification tasks.

\section{Materials and methods}

\subsection{Proteins : Tamapin and Adenylate Kinase}

Here we give a brief description of the two proteins employed in the current work. 

\textit{6d93} is a 31 residue-long mutant of \textit{Tamapin} \cite{Pedarzani_2002}, a toxin of the indian red scorpion. Its outstanding selectivity towards the calcium-activated potassium channels SK2 made it an extremely interesting protein in the field of pharmacology;

\textit{4ake}, the open conformation of \textit{Adenylate Kinase} \cite{muller1996adenylate}. This 214-residue enzyme is responsible for the inter-conversion between Adenosine triphosphate (ATP) and Adenosine diphosphate + Adenosine monophosphate (ADP + AMP) inside the cell.

\begin{figure}[ht]
    \centering
    \includegraphics[width=\columnwidth]{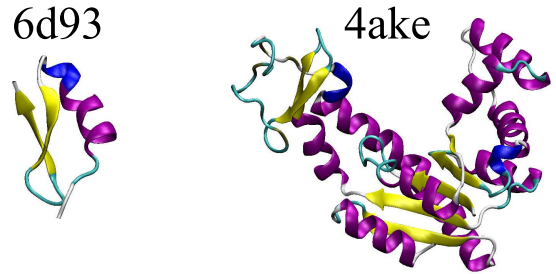}
    \caption{The two protein structures employed in this work, the tamapin mutant (PDB code: 6d93) and the open conformation of Adenylate Kinase (PDB code: 4ake). The former, although small, possesses all the elements of proteins' secondary structures, while the latter is bigger in size and has a much wider structural variability.}
    \label{fig:structures}
\end{figure}

For a concise but more detailed description of these two proteins, please refer to sections II.B and II.D of Ref.~\cite{giulini2020information}.

\subsection{Data representation}

\begin{figure}[t]
    \centering
    \includegraphics[width=0.8\columnwidth]{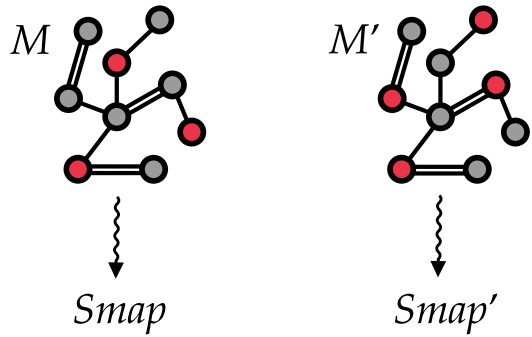}
    \caption{Two different mappings $M$ and $M'$ associated with the same (schematic) protein structure. To train our machine learning model, we treat each protein as a graph where vertices are atoms, and edges are placed among atoms closer than a given threshold. The selected CG sites of the two mappings are marked in red and encoded as a vertex feature. Our goal is to automatically learn to map both mappings to proper values $S_{map}$ and ${S_{map}}'$ of the mapping entropy.}
    \label{fig:mapping}
\end{figure}

To evaluate the efficacy of machine learning methods for graphs in predicting the $S_{map}$ of a mapping, we consider two datasets, namely $6d93_{31}$ and $4ake_{214}$. For each of the proteins described in the previous section we fix the number of CG sites to be equal to the number of amino acids.

\begin{figure*}[t]
\centering
\subfloat{\includegraphics[width=\columnwidth]{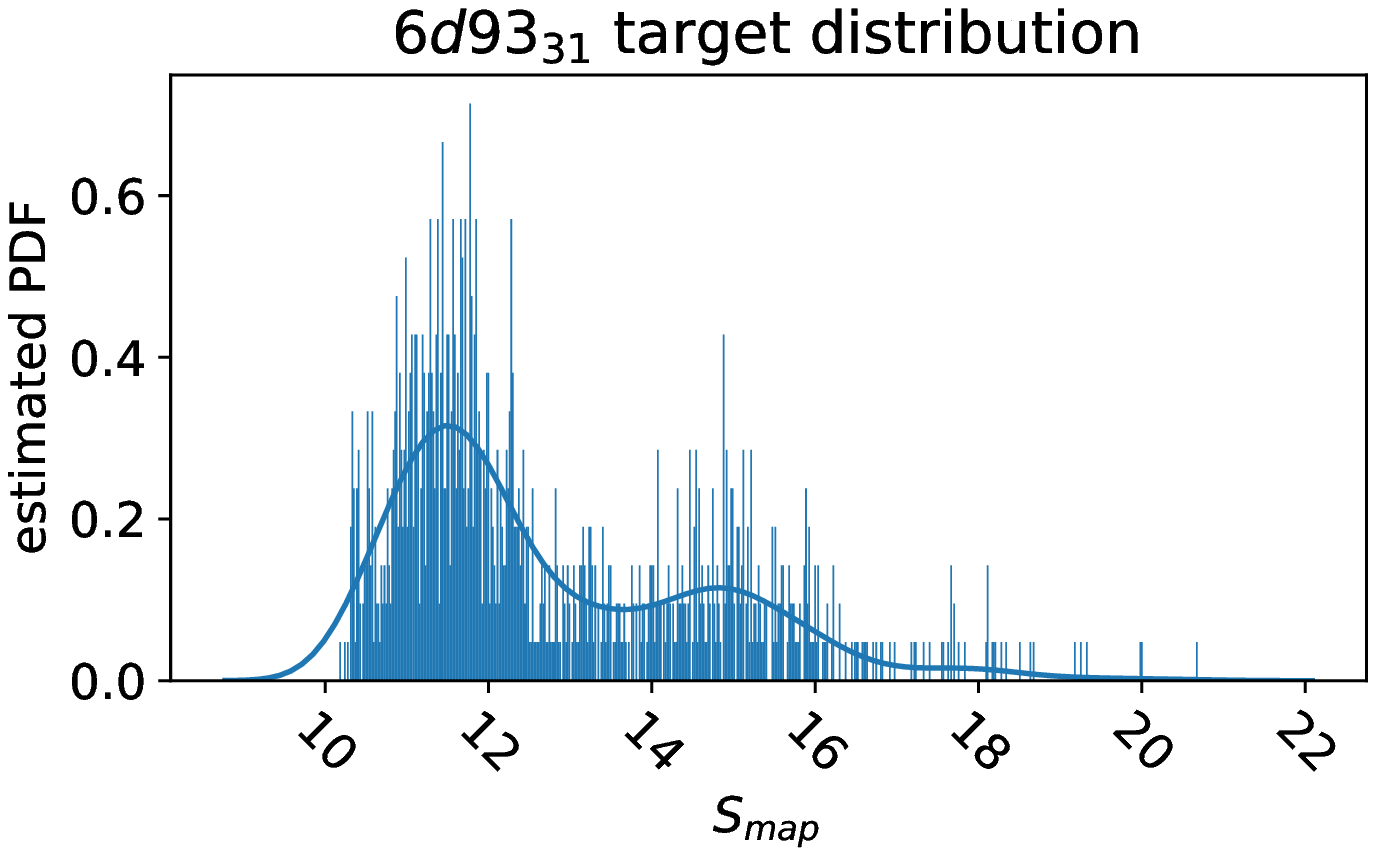}}
\subfloat{\includegraphics[width=\columnwidth]{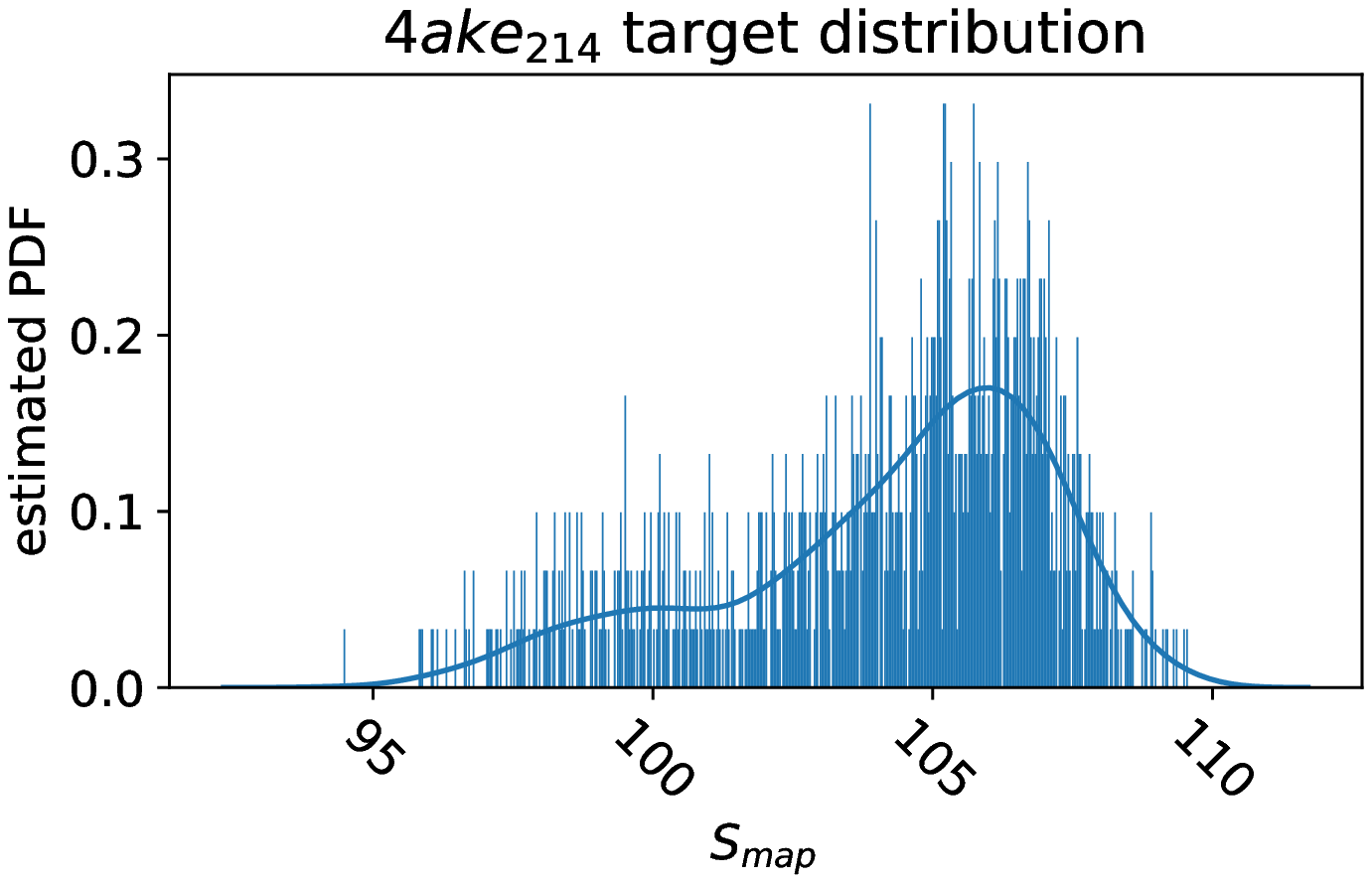}}
\caption{Distributions of target values for both datasets. This is not to be considered the true $S_{map}$ probability density function, since only a fraction of these values (the Gaussian-like curves on the right side of the plots) is associated to a random exploration of the space, while the others are sampled during the optimisation procedure (left region, low $S_{map}$). All values of $S_{map}$ are in $kJ/\text{mol}/K$.}
\label{fig:distributions}
\end{figure*}

Overall, each dataset consists of a single protein structure and many mappings. To be fed into a machine learning model, the structure is represented as a graph where atoms are vertices and edges are inserted whenever two vertices are closer than $1$ nm. The only difference between samples in each dataset is the selection of CG sites, \ie vertices of the graph, which are marked in red in Figure \ref{fig:mapping}. We treat the selection of an atom as a binary input feature attached to the corresponding vertex of the graph. Different selections correspond to different values of $S_{map}$. In addition, we enrich each vertex with 10 features that describe the atomistic physico-chemical properties (see Table \ref{tab:features}); similarly, we consider the inverse atomic distance $e_{uv}$ between vertices $u$ and $v$ as an edge feature. Dataset statistics are summarized in Table \ref{tab:data-statistics}, whereas the distribution of the target values is shown in Figure \ref{fig:distributions}.

Note that different proteins will correspond to graphs with different topology and vertex/edge information, whereas different conformations of the same structure will only differ by their network of edges, since the chemical nature of their atoms remains unaltered.

We build the datasets using the data retrieved in Ref.~\cite{giulini2020information}. Specifically, for both proteins we include 500 randomly selected CG mappings, 48 optimised solutions and 576 intermediate mappings obtained by periodically saving the mappings visited along the first stages of the optimisation procedure.

\begin{table}[ht]
\centering
\begin{tabular}{@{}lll@{}}
\toprule
Feature name &  Value & Description \\ \hline
C & 0/1 & Carbon atom \\
N & 0/1 & Nitrogen atom \\
O & 0/1 & Oxygen atom \\
S & 0/1 & Sulphur atom \\
HPhob & 0/1 & Part of a hydrophobic residue\\
Amph & 0/1 & Part of a amphipathic residue\\
Pol & 0/1 & Part of a polar residue \\
Ch & 0/1 & Part of a charged residue \\
Bkb & 0/1 & Part of the protein backbone \\
Site & 0/1 & Atom selected as a CG site \\

\hline
\end{tabular}
\caption{\label{tab:features}Features used to describe atomistic physico-chemical properties. In this simple model, we only provide the DGN with the chemical nature of the atom and of its residue, together with the flag \textit{Bkb} that specifies if the atom is part of the main polypeptide chain.}
\end{table}

\begin{table}[ht]
    \centering
    \begin{tabular}{lcccc}
    \hline
         Protein    & Graphs & vertices & Edges & Avg Degree \\ \hline
         $6d93_{31}$  & 1124 & 230 & 21474 & 93\\
         $4ake_{214}$ & 1124 & 1656  & 207618 & 125 \\ \hline
    \end{tabular}
    \caption{Basic Statistics of the datasets used.}
    \label{tab:data-statistics}
\end{table}

\subsection{Machine Learning Model}

As mentioned in Section \ref{sec:intro}, the great potential of DGNs lies in the ability of efficiently handling graphs of arbitrary structure and size. This is especially important when we want to approximate complex processes, such as the $S_{map}$ computation on different proteins, in a fraction of the time originally required. The main building block of a DGN is the graph convolution mechanism; in this work, we employ a restricted version of the Gated-GIN model \cite{errica_theoretically_2020} that allows us to consider edge values while keeping the computational burden low. Moreover, we take into account the importance of selected sites by implementing a weighted sum of neighbors:
\begin{align}
     \mathbf{h}_v^{\ell+1} &= MLP^{\ell+1} \Big( \big(1 + \epsilon^{\ell+1} \big)*\mathbf{h}_v^{\ell} + \sum_{u \in \mathcal{N}_v} \mathbf{h}_u^\ell*e_{uv} \Big) \\
     \hat{S}_{map} &= \mathbf{w}_{out}^T \Big(ReLU \Big(\sum_{u \in \mathcal{V}^s_g} W^T[\mathbf{h}_u^1,\dots,\mathbf{h}_u^L]*w_s \nonumber \\
    & + \sum_{u \in \mathcal{V}_g \setminus \mathcal{V}^s_g} W^T[\mathbf{h}_u^1,\dots,\mathbf{h}_u^L]*w_n \Big) \Big)
\end{align}
where $\mathcal{N}_v$ represents the neighborhood of $v$, $*$ is scalar multiplication, $e_{uv}$ is a scalar edge feature holding the inverse atomic distance between two atoms $u$ and $v$, $\epsilon$, $W$, $\mathbf{w}_{out}$, $w_n$ and $w_s$ are adaptive weights related to unselected and selected CG sites, respectively, $MLP$ is a multi-layer perceptron of two linear layers interleaved by a Rectifier linear unit (ReLU) activation function \cite{glorot_deep_2011}, and square brackets denote concatenation of the different vertex embeddings computed at different layers. Depending on whether a vertex has been selected as a site, \ie $u \in  \mathcal{V}^s_g$, or not, \ie $u \in \mathcal{V}_g \setminus \mathcal{V}^s_g$, different weights are learned. This way, the model is able to perform a \quotes{site-aware} aggregation of vertex embeddings before predicting the mapping entropy of the whole graph.
\section{Results and discussion}

To assess the performance of the model on a single protein, we first split the dataset into training, validation and test realizations following an 80\%/10\%/10\% hold-out strategy. We trained and assessed the model on each dataset separately. We applied early stopping \cite{prechelt_early_1998} to select the training epoch with the best validation score, and the chosen model was evaluated on the unseen test set. The evaluation metric for our regression problem is the coefficient of determination (or R$^2$-score).

Since training can be costly depending on the protein, in these preliminary experiments we tried a single configuration with $5$ graph convolutional layers where each layer uses $64$ hidden units (\ie the length of $\mathbf{h}_v^\ell$) inside the $MLP$. $W$ is the weight matrix of a linear layer with $16$ output units and $\mathbf{w}_{out}$ is another weight vector of a linear layer that maps the input to a single scalar value. The loss objective is the Least Absolute Error. The optimization algorithm is Adam \cite{kingma_adam_2015} with a learning rate of $0.001$ and no regularization. We trained for a maximum of $10000$ epochs with early stopping patience of $1000$ epochs and mini-batch size $8$. We accelerated training using a Tesla V100 GPU with 16GB of memory.


\begin{table*}[ht]
    \centering
    \small
    \begin{tabular}{lcccccc}
    \hline
         Protein    & TR MAE & TR Score & VL MAE & VL Score & TE MAE & TE Score \\ \hline
         $6d93_{31}$  & 0.11 & 0.99 & 0.41 & 0.94 & 0.56 & 0.85 \\
         $4ake_{214}$ & 0.90 & 0.89  & 1.05 & 0.88 & 1.22 & 0.84 \\ \hline
    \end{tabular}
    \caption{Result on the training (TR), validation (VL) and test (TE) sets for both datasets. The mean average error (MAE) is rounded to its closest integer.}
    \label{tab:results}
\end{table*}

\begin{figure*}[ht]
\centering
\subfloat{\includegraphics[width=\columnwidth]{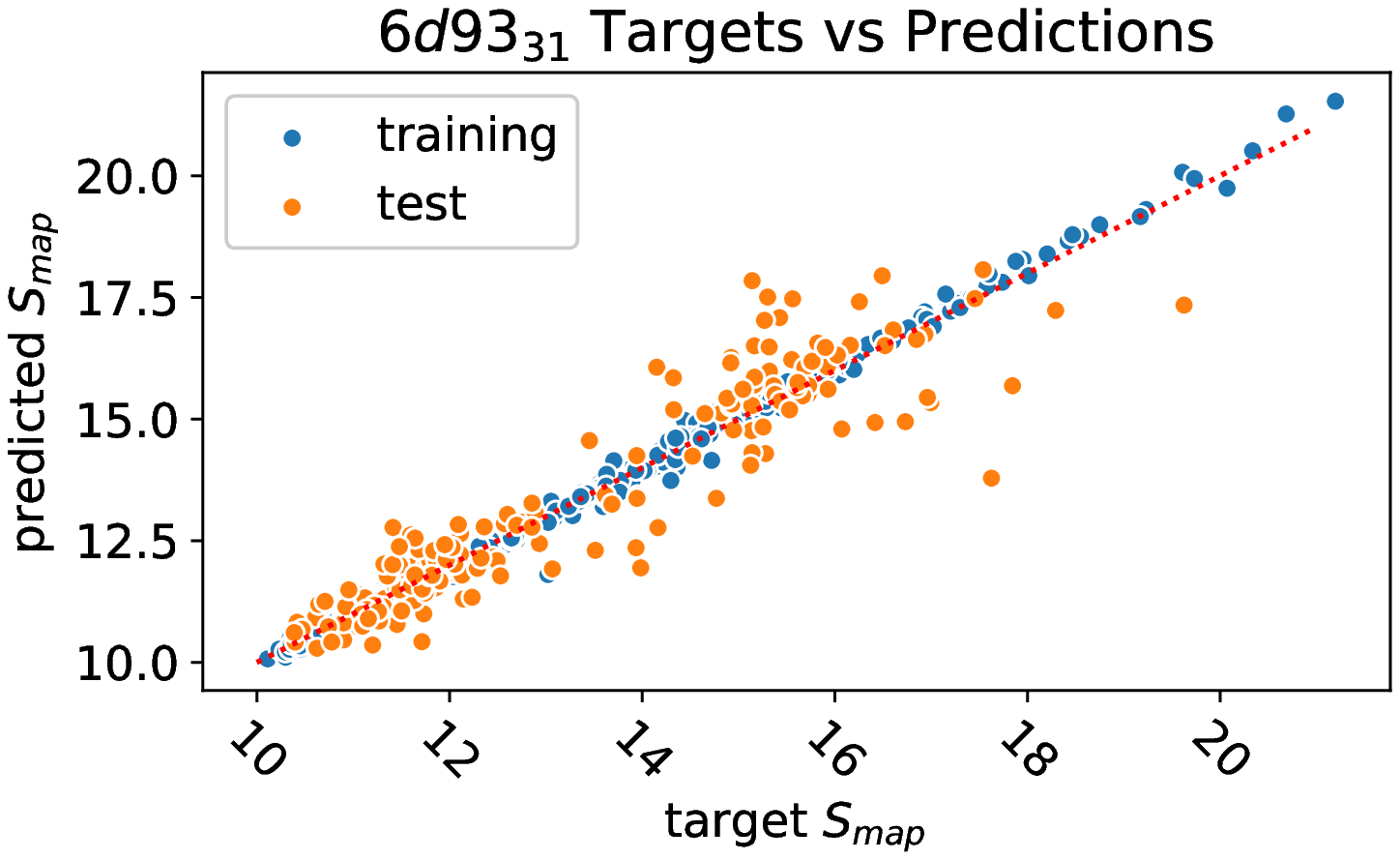}}
\subfloat{\includegraphics[width=\columnwidth]{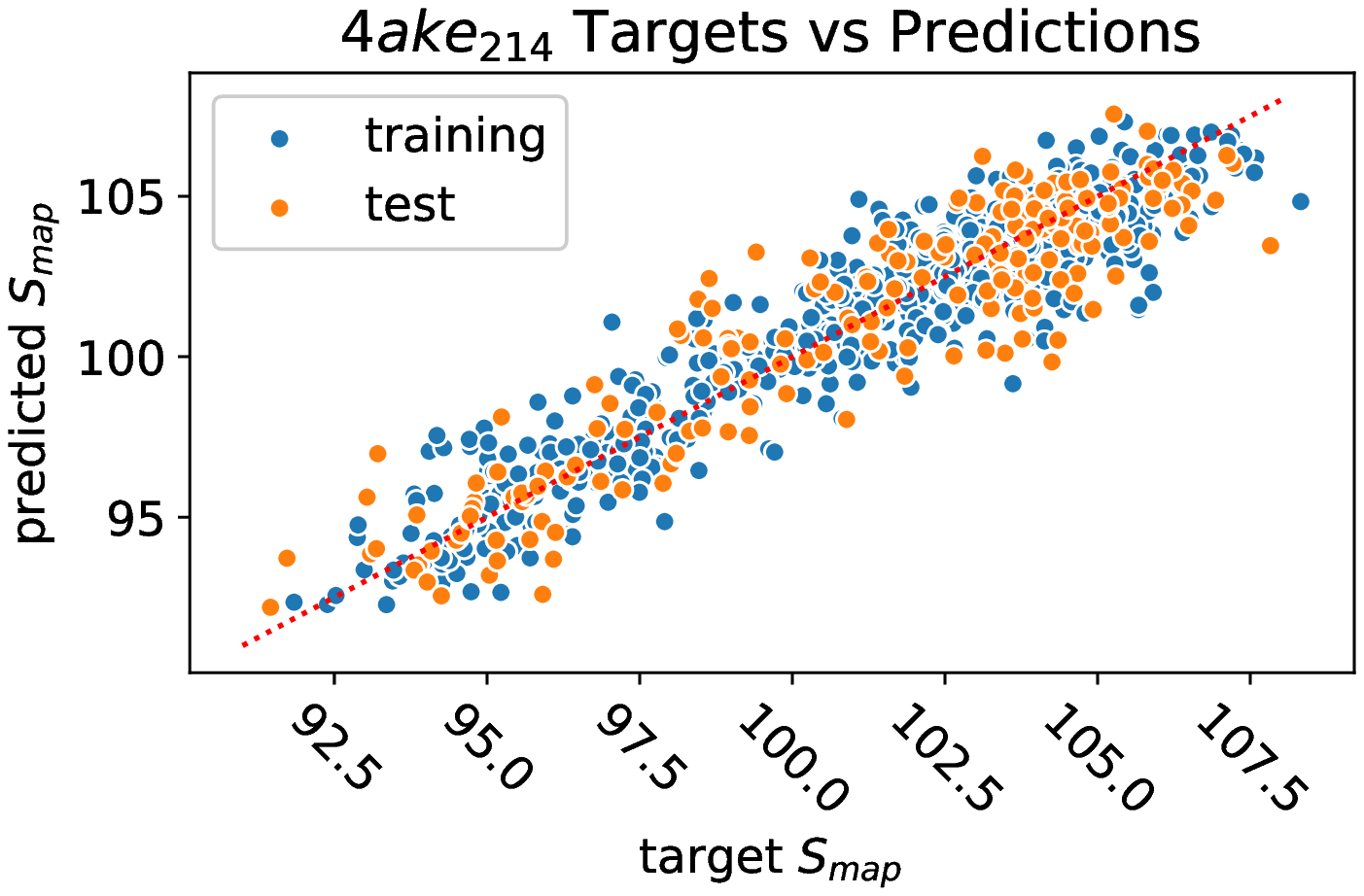}}
\caption{Plot of target values against predictions of all samples. A perfect prediction is represented by the dotted diagonal line. Training samples are in blue, while test samples are in orange. All values of $S_{map}$ are in $kJ/\text{mol}/K$.}
\label{fig:predvstest}
\end{figure*}

\begin{figure*}[ht]
\centering
\subfloat{\includegraphics[width=\columnwidth]{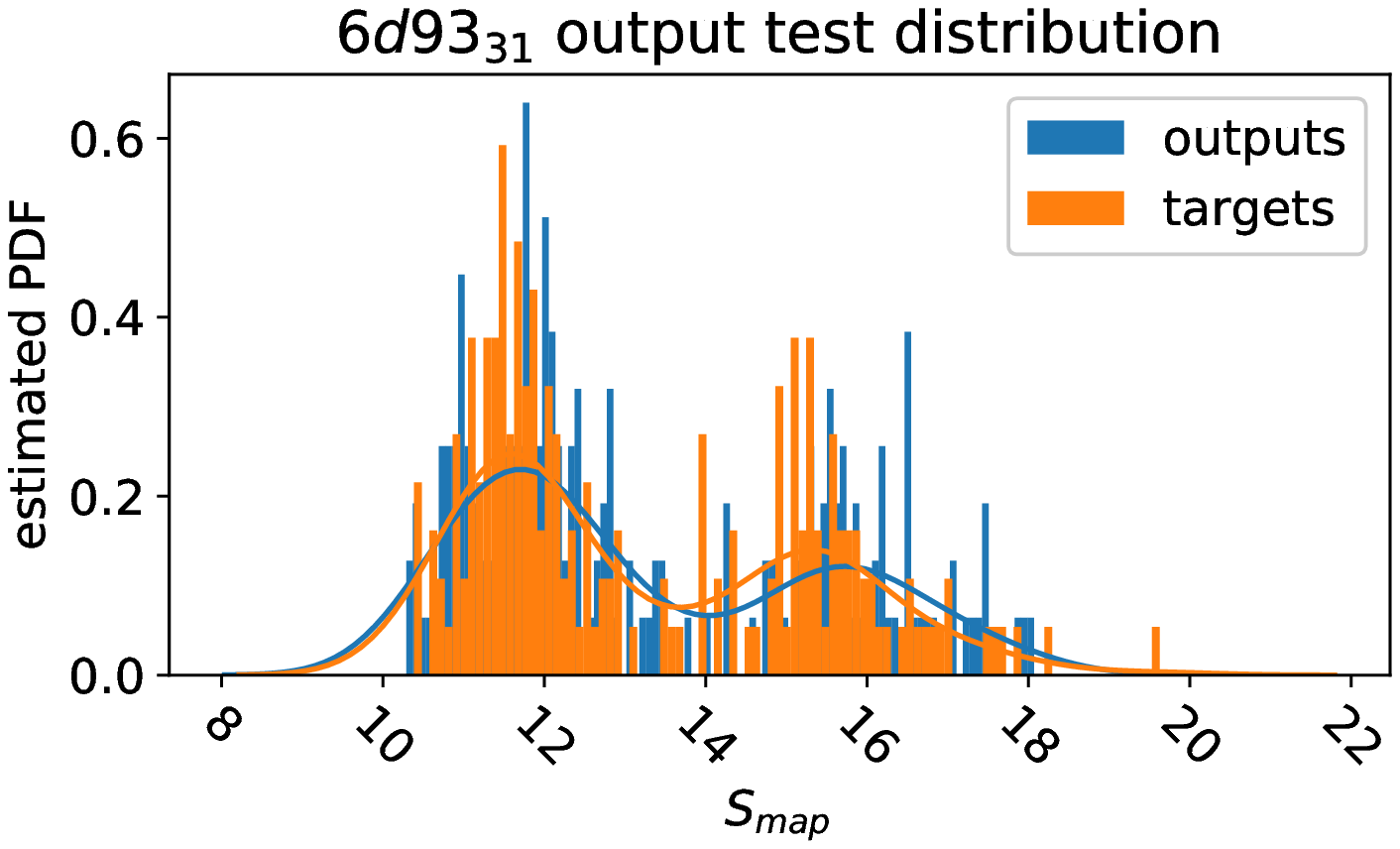}}
\subfloat{\includegraphics[width=\columnwidth]{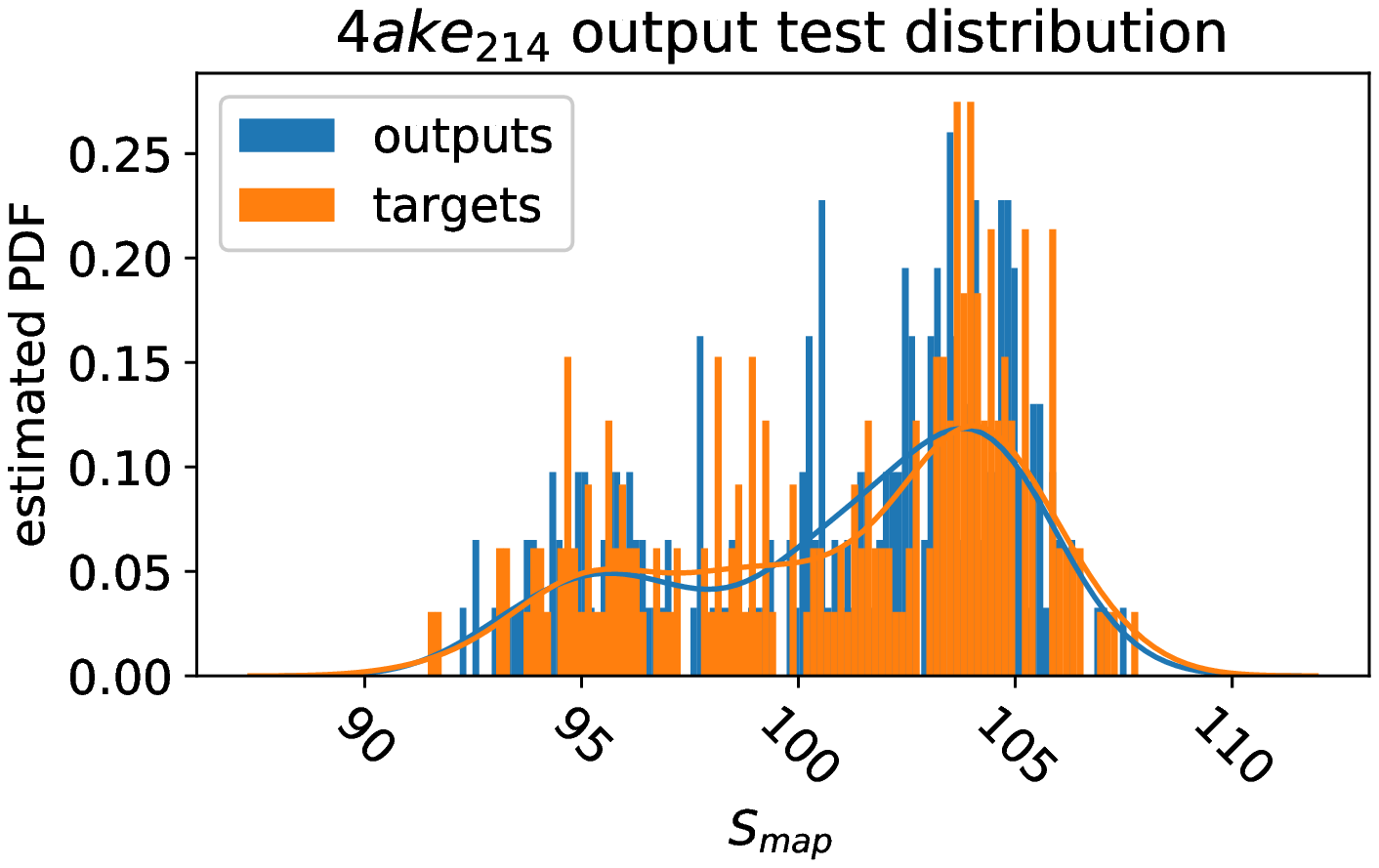}}

\caption{Output distribution on test samples for both datasets. All values of $S_{map}$ are in $kJ/\text{mol}/K$.}
\label{fig:outdist}
\end{figure*}

Table \ref{tab:results} reports the R$^2$-score for the $6d93_{31}$ and $4ake_{214}$ datasets in both training and test. As we observe, the model can fit the training set, and it has very good performances on the test set. In Fig. \ref{fig:predvstest}, we show how predicted values for training (blue) and test (orange) samples differ from the ground truth. Ideally, a perfect prediction corresponds to the point being on the diagonal dotted line. For completeness, Fig. \ref{fig:outdist} compares the distribution of the predicted values (blue) with that of the ground truth (orange), and we observe that the distribution of the outputs produced by the model closely follows the target one.

Table \ref{tab:time-inference} shows a time comparison between the algorithm that computes the $S_{map}$ value and the inference of the model. Generally, the cost of computing the mapping entropy is drastically reduced by $2-5$ orders of magnitude depending on the hardware used. This means that, in principle, we can explore many more mappings in order to characterise the mapping entropy landscape and, in particular, its minima.

\begin{table*}[ht]
\centering
\begin{tabular}{@{}llll@{}}
\toprule
Protein &  Single measure & Inference GPU (CPU) & Time Ratio GPU (CPU)  \\ \hline
$6d93_{31}$ & $\simeq 2.1 $ mins  & $\simeq 0.9 (98.7)$ ms & $\simeq 140000\times (1276\times)$\\
$4ake_{214}$ & $\simeq 8.0 $ mins & $\simeq 4.8 (1103.2) $ ms & $\simeq 100000\times (435\times)$\\  \hline
\end{tabular}
\caption{\label{tab:time-inference} Comparison between the time required to compute $S_{map}$ of a single mapping and the inference time of the model (both on CPU and GPU). We obtain a drastic improvement that allows a wider exploration of the $S_{map}$ space.}
\end{table*}

\section{Conclusions and perspectives}

Molecular dynamics simulations constitute the core of the majority of research studies in the field of computational biophysics. From protein folding to free energy calculations, an all-atom trajectory of a biomolecule gives access to a vast amount of data about the system, from which relevant information about the system's properties, behaviour, and biological function is extracted through \textit{a posteriori} analysis. This information retrieval can be almost immediate and intuitive to observe (even by naked eye) and quantify in terms of few simple parameters (e.g. the process of ligand binding can be seen in a graphical rendering of the trajectory, and made quantitative in terms of the distance between ligand and protein); much more frequently, though, it is a lengthy and non-trivial task.

The mapping entropy method, introduced in Eq. \ref{eq:smap_general} and here briefly summarised, aims at performing in an automated, unsupervised manner this identification of relevant features of a system--namely, the subset of a molecule's atoms that retain the largest amount possible of information about its behaviour. This scheme relies on the computation of $S_{map}$, that is, a measure of the dissimilarity between the original system configurations' probability density and the one marginalised over the discarded atoms. This quantity is employed as a cost function and minimised over the possible reduced representations, or mappings.

The extent of the mapping space, however, makes a random search practically useless and an exhaustive counting simply impossible; hence, an optimisation procedure is required to identify the simplified description of the molecule that entails the largest amount of information about the system. Unfortunately, this procedure nonetheless implies the calculation of $S_{map}$ over a very large number of tentative mappings, making the optimisation, albeit possible, computationally intensive and time consuming.

In this work, we have tackled the problem of speeding up the $S_{map}$ calculation procedure by means of deep learning algorithms. In particular, we have shown that Deep Graph Networks are capable of inferring the value of the mapping entropy when provided with a schematic, graph-based representation of the protein and a tentative mapping. The method's accuracy is tested on two proteins of very different size, tamapin (31 residues) and adenylate kinase (214 residues), with a test score of $0.85$ and $0.84$, respectively. 
These rather promising results have been obtained in a computing time that is up to five orders of magnitude shorter than the standard algorithm.

The presented strategy can hold the key for an exhaustive exploration of the CG mappings of a biomolecule. In fact, even if the training process has to be carried out for each protein anew, the trained DGN algorithm can be employed to extend the characterisation of the mapping entropy landscape of the system with impressive accuracy in a fraction of the time.

The natural next step would be to apply the knowledge acquired by the model on different protein structures, so that the network can predict values of $S_{map}$ even in the absence of an MD simulation. As of now, however, it is difficult to assess if the information extracted from the training over a given protein's trajectory can be fruitfully employed to determine the mapping entropy of another, just feeding the structure of the latter as input. In other words, obtaining a transfer effect by the learning model may not be straightforward, and additional information will be needed to achieve it.

In conclusion, we point out that the proposed approach is completely general, in that the specific nature and properties of the mapping entropy played no special role in the construction of the deep learning scheme; furthermore, the DGN formalism enables one to input graphs of variable size and shape, relaxing the limitations present in other kinds of DL architectures \cite{giulini2019deep}. This method can thus be transferred to other problems where different selections of a subset of the molecule's atoms give rise to different values of a given observable (see \eg \cite{potestio_jctc}), and pave the way for a drastic speedup in computer-aided computational studies in the fields of molecular biology, soft matter, and material science.

\begin{acknowledgments}
This project has received funding from the European Research Council (ERC) under the European Union's Horizon 2020 research and innovation programme (grant agreement No 758588).
\end{acknowledgments}

\bibliographystyle{ieeetr}
\bibliography{main}

\end{document}